\documentclass[12pt]{iopart}

\usepackage{iopams}
\usepackage{graphicx}
\usepackage{dcolumn}
\usepackage{bm}

\begin{document}

\title[YbF fountain to measure the electron edm]{Design for a fountain of YbF molecules to measure the electron's electric dipole moment}

\author{M~R~Tarbutt, B~E~Sauer, J~J~Hudson, and E~A~Hinds }

\address{
Centre for Cold Matter, Blackett Laboratory, Imperial College London, Prince Consort Road, London SW7 2AZ, United Kingdom
}
\eads{\mailto{ed.hinds@imperial.ac.uk}, \mailto{m.tarbutt@imperial.ac.uk}}

\begin{abstract}
We propose an experiment to measure the electric dipole moment of the electron using ultracold YbF molecules. The molecules are produced as a thermal beam by a cryogenic buffer gas source, and brought to rest in an optical molasses that cools them to the Doppler limit or below. The molecular cloud is then thrown upward to form a fountain in which the electric dipole moment of the electron is measured.
\end{abstract}

\maketitle

\section{\label{sec:Introduction} Introduction}

\begin{figure}
\begin{center}
\includegraphics[scale=0.36]{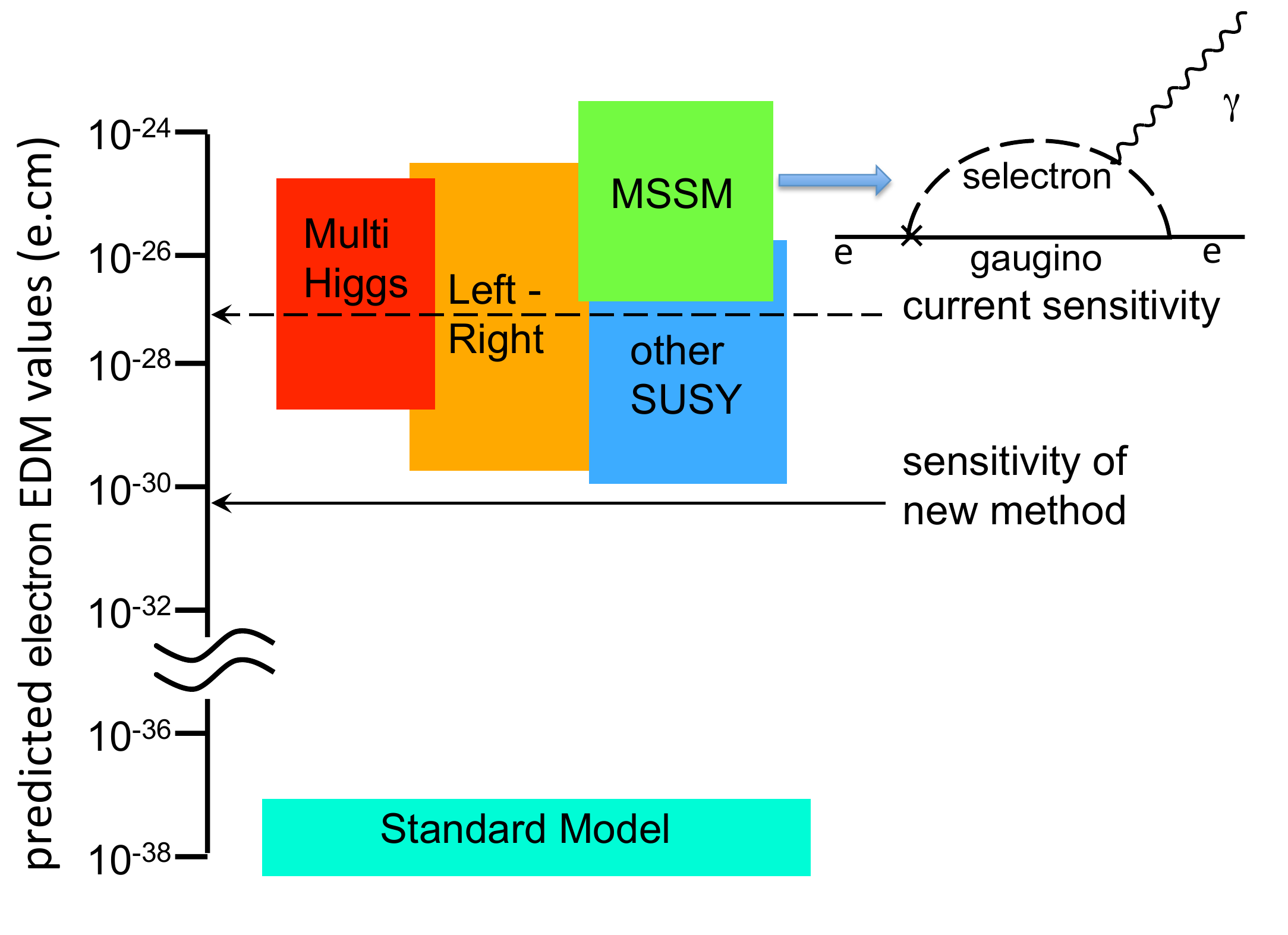}
\caption{\label{fig:limits}Coloured boxes: range of predicted values for the EDM. Dashed line: current upper limit on electron EDM \cite{Hudson11}.
 Solid line: sensitivity of proposed new method. Top right diagram: example of a simple supersymmetry 1-loop radiative correction, generating an electron EDM. A cross marks the CP- violating vertex.}
\end{center}
\end{figure}

The standard model of elementary particle physics predicts that the electric dipole moment (EDM) of the electron is exceedingly small \cite{Lebedev02}, $d_e\simeq10^{-38}$e.cm, as illustrated in figure \ref{fig:limits}. In all local, Lorentz-invariant field theories a permanent EDM requires CP symmetry to be violated. This happens in the standard model through the Yukawa couplings in the quark sector \cite{CKM73}, with the first non-vanishing result appearing at the four-loop level \cite{Khriplovich91}. However, the standard model is thought to be incomplete, one reason being that it does not have enough CP violation to explain the excess of matter over antimatter in our universe, as first discussed by Sakharov \cite{Sakharov67}. It would seem that new forces are waiting to be discovered, involving new particles with masses probably near the electroweak scale of 0.2 TeV up to a few TeV. This motivates many current experiments in particle physics.

The electron is sensitive to these new CP-violating forces because they are expected to induce an EDM, even at the level of a simple one-loop radiative correction, such as the one shown in the upper right of figure \ref{fig:limits}. Here the natural size of the
EDM is \cite{Pospelov05} $\frac{\alpha}{\pi}\frac{m_e}{M_{s}^{2}}\frac{e \hbar}{c}\sin(\theta_{CP})$, where $\alpha$ is the fine structure constant, $m_e$ is the mass of the electron, $M_s$ is the mass of the new particle, here the selectron, and $\theta_{CP}$ is the CP violating phase of the coupling. Since there is no reason to assume CP conservation, let us take $\theta_{CP}\simeq 1$ , then $M_{s}=1$TeV$/c^2$ gives $d_{e}\simeq 2\times10^{-26}$e.cm. Similarly, $M_{s}=100$TeV$/c^2$ gives $d_{e}\simeq 2\times10^{-30}$e.cm.  In this example the virtual selectron and gaugino are supersymmetric particles, but most other models similarly predict EDM values \cite{Khriplovich97,Commins99} in the range  $10^{-26}-10^{-30}$e.cm, as shown in figure \ref{fig:limits}. If there is new particle physics in the energy range 200 GeV--100 TeV, an electron EDM sensitivity of $10^{-30}$e.cm is very likely to detect it.

Experiments going back to the 1960s \cite{Sandars64} have sought to detect the electron EDM using beams of heavy atoms. These have the virtue that the interaction energy $\eta d_{e} E$ between the electron EDM $d_e$ and the applied electric field $E$ is enhanced by the relativistic factor $\eta$, which can be large, as discovered by Schiff \cite{Schiff63} and Sandars \cite{Sandars65}. The atomic beam experiments ended in 2002 with a measurement on Tl, a complex tour de force that gave the limit \cite{Regan02} $|d_{e}|\le 1.6\times10^{-27}$e.cm. In order to improve on this, a new technique has been developed using polar molecules \cite{Hinds97}, which offer two great advantages. (i) Molecules are more sensitive to $d_e$ than atoms. For example, in a large laboratory field YbF molecules are 200 times more sensitive than Tl atoms \cite{enhancement}. (ii) Molecules can be immune to the particular systematic error (generated by the motional magnetic field $E\times v/c^2$ ) that limited the Tl experiment \cite{Hudson02}. The recent electron EDM measurement using this new method \cite{Hudson11} has given a slightly improved upper limit of $|d_{e}|\le 1.05\times10^{-27}$e.cm, shown by the dashed line in figure \ref{fig:limits}. Together, these measurements indicate that the simple supersymmetric model illustrated in figure \ref{fig:limits} cannot be right: either the CP violation is suppressed for some reason or the mass of the new particle is considerably greater than 1 TeV. While it is possible to construct supersymmetric theories where some particular EDM is accidentally small, it does not seem possible in a minimal supersymmetric model to keep the electron and neutron EDM both small enough \cite{Pospelov05,Commins10}. One is led to the conclusion that either the superpartner masses are rather heavy or there is some more complex physics on the electroweak scale that breaks supersymmetry but manages to suppress CP-violation. This is an important open question at a forefront of fundamental physics that will be elucidated by a more sensitive search for the electron EDM.

To go significantly beyond the current level of EDM sensitivity, one needs a fundamental improvement in the method, which is what we propose here. By laser-cooling the molecules, it will be possible to replace the molecular beam by a molecular fountain, where each molecule can precess coherently for almost one second, rather than the current one millisecond. This will allow detection of an EDM as small as $1\times10^{-30}$e.cm, corresponding to CP-violating physics at energies up to 100 TeV. With this sensitivity one can surely hope to uncover some of the new physics underlying the important issue of CP-violation in the early universe.

Quite apart from this application in elementary particle physics, there is strong interest in cooling molecules to microKelvin temperatures for possible use in quantum information processing \cite{DeMille02,Andre06}, simulation of quantum many-body systems \cite{Goral02,Micheli06}, ultracold chemistry \cite{Balakrishnan01,Krems08}, nuclear physics \cite{DeMille08}, and cosmology \cite{Flambaum07}.

One approach is to photo-associate or magneto-associate atoms that are first separately cooled to ultra-low temperatures. This is proving to be a powerful tool for studying the physics of ultracold molecular gases \cite{Miranda11,Danzl10}. However, the method is limited to a few alkali/alkali and alkali/alkaline-earth diatomics and is therefore not suitable for many of the applications listed above that require other types of molecule. By contrast, a huge range of molecular species can be produced by supersonic expansion or by cryogenic buffer gas cooling \cite{Krems09}. These methods produce molecular beams whose temperatures are typically a few Kelvin. This is not cold enough for the proposed molecular fountain which requires a temperature in the microKelvin range. For this reason, it is important to find a cooling method that bridges the gap between a few K and 100\,$\mu$K.

One possibility for bridging this gap is to trap the molecules together with ultracold atoms and allow them to cool sympathetically as a result of their collisions with the atoms \cite{Carvalho99,Lara06,Soldan09,Wallis09,Tokunaga11}. It is possible that this method will yield very cold, dense molecular samples if we can find cases where the elastic collision cross section is large enough and the losses are small enough. However, the number of molecules may well be  inadequate for EDM measurements, which require low statistical noise and must therefore detect many molecules.

Here, we propose the promising alternative of using laser cooling, a method already well established for atoms. Most molecules cannot be laser-cooled because they decay after a few spontaneous emissions into a higher vibrational state that does not interact with the cooling light. However, there is a family of molecules where this difficulty is not too severe \cite{DiRosa04}. De Mille's group at Yale has recently demonstrated laser cooling of SrF molecules \cite{Shuman10,Barry11}. In the case of YbF, our group has measured the Franck-Condon factors \cite{Zhuang11}, the ratios that characterise the change of vibrational state, and find them to be suitable for laser cooling.

\begin{figure}
\begin{center}
\includegraphics[scale=0.7]{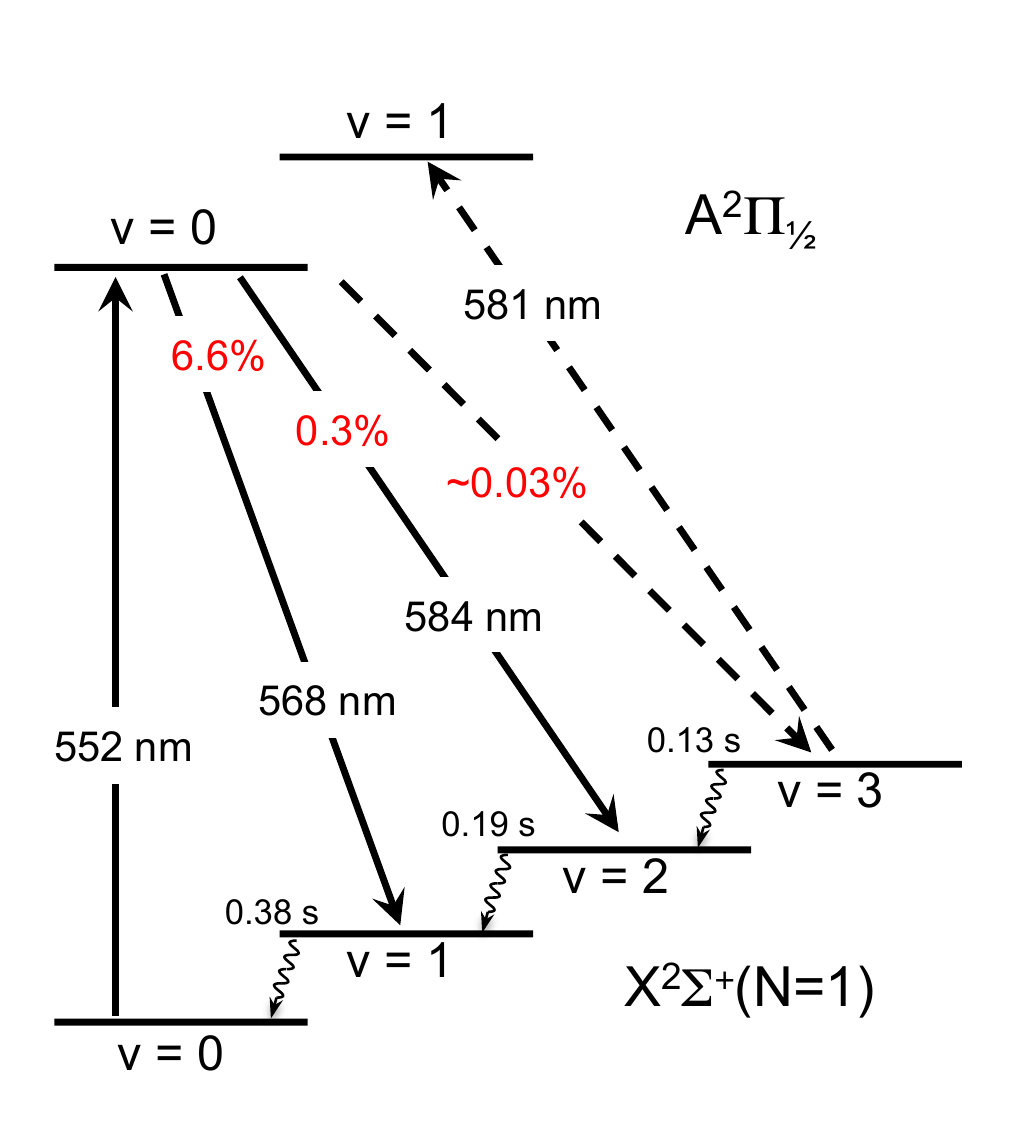}
\caption{\label{fig:levels}
Scheme for laser cooling YbF. The cooling is due to multiple scattering of 552\,nm light on the A$^{2}\Pi_{1/2}$ - X$^{2}\Sigma^{+} (v = 0 - 0)$ transition. Molecules that decay to $v = 1$ and $v = 2$ are quickly repumped to $v = 0$ using lasers at 568\,nm and 584\,nm. Molecules that fall into the $v = 3$ state are cold and require only weak repumping. The branching ratios to the various vibrational levels, and the lifetimes for spontaneous vibrational decay in the X state, are also shown.
}
\end{center}
\end{figure}

Figure \ref{fig:levels} shows the relevant energy levels of YbF. The lowest odd-parity state X$^{2}\Sigma^{+}(v=0,N=1)$ is excited on the P(1) line to the even parity A$^{2}\Pi_{1/2}(v=0)$ state by near-resonant light of wavelength $\lambda_{00}=552$\,nm. This can only decay to X$^2\Sigma^+$ states having $N=1$ because the next odd-parity ground state, $N=3$, is forbidden by the angular momentum selection rules for electric dipole radiation. Each scattered photon imparts a momentum $h/\lambda_{00}$ to the molecule ($h$ being Planck's
constant), corresponding to a velocity change of 3.7 mm/s. By comparison, YbF molecules in thermal equilibrium at a temperature of $T = 3.6$\,K have an rms velocity along a given direction of $\sqrt{k_{B}T/M}= 12$\,m/s, where $k_B$ is the Boltzman constant and $M$ is the mass of one molecule. Thus, molecules pre-cooled to this temperature by a helium buffer gas can be brought to rest by scattering a few thousand photons. Six beams propagating along the Cartesian directions $\pm\hat{x},\pm\hat{y},\pm\hat{z}$ are detuned by approximately 5 MHz to the red of resonance in order to form an optical molasses \cite{Metcalf99}. This cools the molecules to a temperature of order $\hbar \Gamma/(2k_B)=140\mu$\,K, where $\Gamma$ is the spontaneous decay rate of the A state. There is a $6.6\%$($0.3\%$) probability that the electronic decay will excite the $v=1$($v=2$)  vibrational state\,\cite{Zhuang11}. These molecules are returned to the cooling cycle by repump lasers at $568$\,nm ($584$\,nm). We show below that a 3.6 K molecular beam can be cooled to form a molecular fountain and that this will improve the experimental EDM sensitivity by a very large factor, as indicated by the solid arrow in figure \ref{fig:limits}.

\begin{figure}
\begin{center}
\includegraphics[scale=0.5]{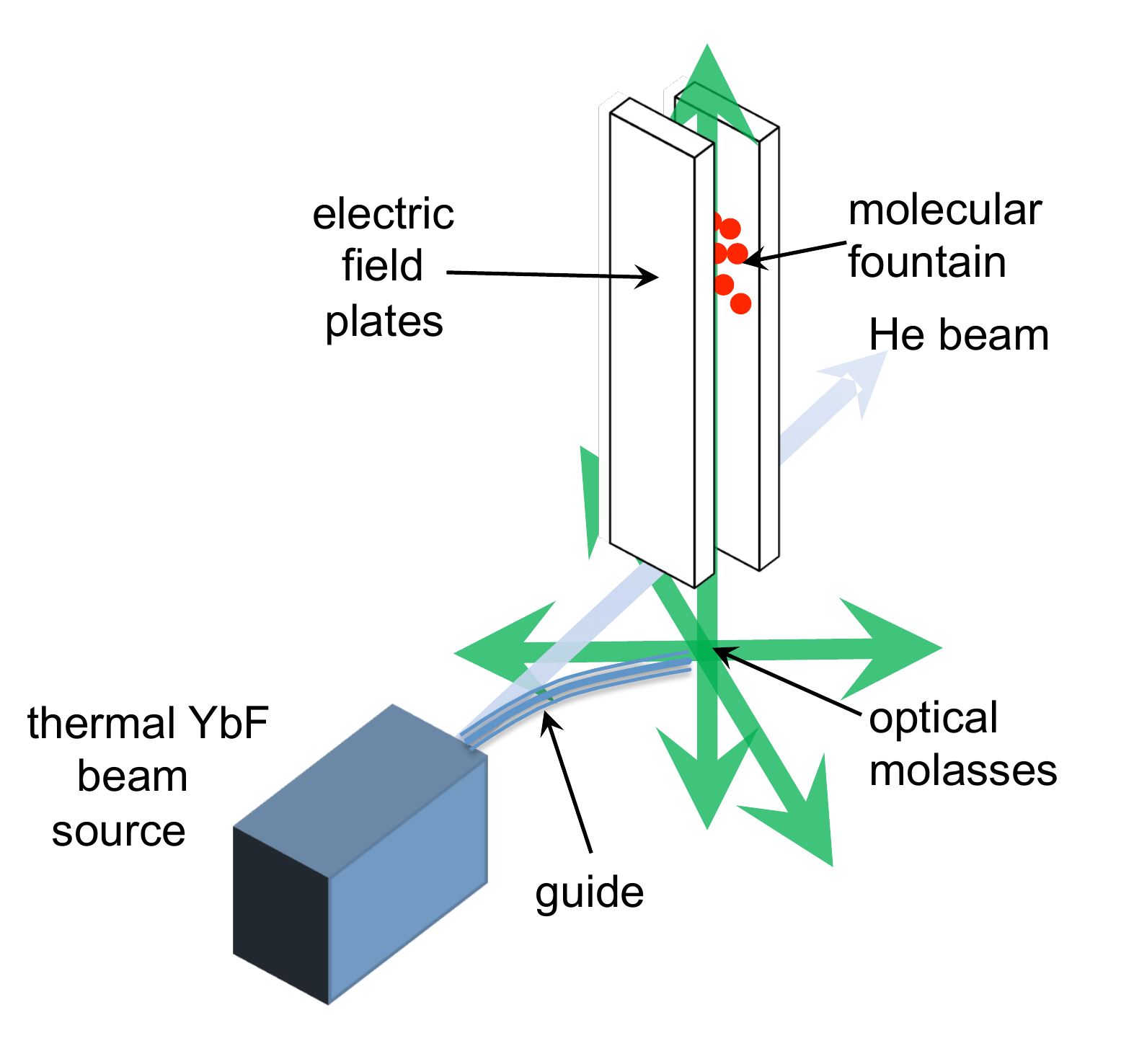}
\caption{\label{fig:overview}The main components of the proposed experiment.
}
\end{center}
\end{figure}

Figure \ref{fig:overview} illustrates the main components of the proposed experiment. The YbF molecules are first produced as a pulsed 3.6\,K beam through thermalisation with a cryogenic helium buffer gas. They are then steered away from the helium beam and into a region of lower background gas pressure by a magnetic guide. On exiting from the guide each cloud of molecules encounters a region of optical molasses, where it is stopped and cooled to ultralow temperature. The cold molecules are then thrown up as a fountain into the space between two electric field plates. There the electron spin is first polarised, then allowed to precess in the electric field. Finally, the spin direction is read out and the electron EDM is deduced from the amount of precession. In the following sections we describe each of these steps in more detail.

\section{\label{sec:Beam} Thermal beam}

\begin{figure}
\begin{center}
\includegraphics[scale=0.5]{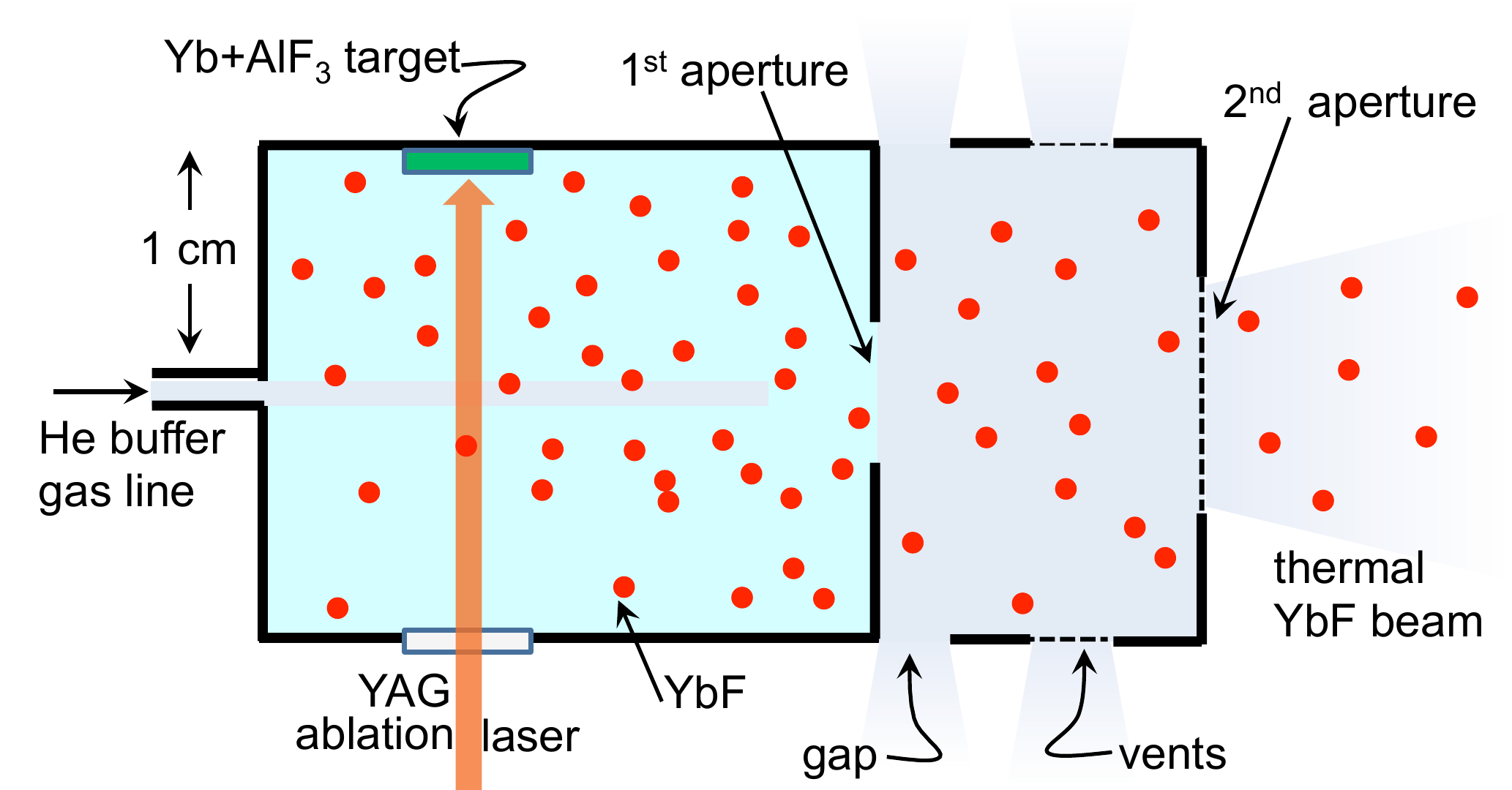}
\caption{\label{fig:source}Cryogenic, thermal molecular beam source design, taken from \cite{Lu11}.}
\end{center}
\end{figure}

The YbF molecules will be prepared as a thermal beam at a temperature of 3.6 K, using the double cryogenic cell design developed by Doyle's group \cite{Lu11} and illustrated in figure \ref{fig:source}. Helium gas enters a first cryogenic cell, where a target containing Yb metal and AlF$_3$ is illuminated by pulses of infrared light from a YAG laser. The materials ablated from the target react to produce a cloud of YbF molecules. These thermalise with the cold helium gas but emerge through the first aperture with a boosted centre-of-mass velocity due to the flow of the helium. The resulting velocity distribution is well described by  $f(u)\propto u^{3}\exp{(-M(u-u_0)^2/(k_{B}T}))$, where $u_{0}$ is typically $100-200$\,m/s. The second cell, where the pressure and helium flow velocity are much lower, reduces $u_{0}$ close to zero. The collisions in this cell reduce the total number of molecules in the beam, but greatly increase the number of slow molecules as shown experimentally in \cite{Lu11}. This is important because the laser cooling cannot bring faster molecules to rest. In our own group, we have studied the production of YbF in a single cryogenic buffer gas cell \cite{Skoff09,Skoff11}. We know from these experiments that we can produce $1\times 10^{13}$ X-state molecules per pulse in the cell \cite{Skoff11} at the He density that optimises thermal beam production. We have extracted a YbF beam from such a cell \cite{SkoffThesis}, but have not tried making a thermal beam using a two-cell source of this type. Reference \cite{Lu11} reports an extraction efficiency of 0.5\% for CaH, and we see no reason why the same should not apply to YbF. With this assumption, we expect a thermal beam of $5\times10^{10}$ X- state molecules per pulse at a temperature of 3.6 K, of which 24\% (according to the Boltzmann distribution) will be in the state $X^2\Sigma^{+}(v=0,N=1)$ that is to be laser cooled. These numbers are recorded for later reference in Table\,\ref{tab:accounts}.

\begin{table}[b!]
\caption{
\label{tab:accounts}
Factors that determine the number of molecules available for the EDM fountain experiment.
}
\begin{center}
\begin{tabular}{rc}
\hline
\vspace{-6 pt}\\
Number of X-state molecules/pulse in cell&$\;1\times10^{13}\;$\\
Fraction in thermal beam&$\;5\times10^{-3}\;$\\
Fraction having $(v=0, N=1)$&$\;2.4\times10^{-1}\;$\\
Guide transmission &$\;6.5\times10^{-2}\;$\\
Fraction cooled by molasses &$\;7.6\times10^{-3}\;$\\
Fraction returning from fountain &$\;7.5\times10^{-2}\;$\\
\hline
\vspace{-8 pt}\\
Total number of cold molecules/pulse&$\;4.4\times10^{5}\;$\\
\vspace{-8 pt}\\
\hline
\end{tabular}
\end{center}
\end{table}

\section{\label{sec:Guide} Magnetic Guide}

A 20\,cm long magnetic guide collects the molecules that emerge from this source and bends them out of the helium beam. The bend radius is large enough that only a small fraction of the very fastest molecules are lost, and these cannot be captured by the molasses anyway. Magnetic guiding from a cryogenic effusive beam has previously been demonstrated using O$_{2}$ molecules \cite{Patterson07}. We propose the same type of guide: a magnetic octupole built from Nd:Fe:B magnets. The eight poles are arranged with their faces touching a circle of radius $5\,$mm, and with gaps of $1\,$mm between their corners. According to a numerical model of the magnetic field, this trap has a depth of $0.6\,$T and therefore confines weak-field-seeking molecules whose transverse velocity is below $5.9\,$m/s, amounting to $13\%$ of the Boltzmann distribution. Since half the molecules are in weak-field-seeking states, the transmission of the guide is expected to be $6.5\%$. There should be no significant loss of molecules from collisions with the helium because the cloud coming out of the cell is thermal, indicating that the molecules are decoupled from the much faster helium beam. Spin flip losses are also negligible.

The magnetic field needs to decrease greatly outside the guide, because the laser cooling will not work well if the field exceeds $1\,$mT. It must also decrease rapidly because the cloud of molecules expands as soon as it leaves the guide. A steel annulus ($5\,$mm inner radius and $2\,$mm thick) attached to the end of the guide provides a good flux return path between magnets and forces the field to drop to an acceptable level only $4\,$mm from the end of the permanent magnets.

\section{\label{sec:Cooling} Laser Cooling}

\begin{table}[b!]
\caption{
\label{tab:hfs}
Hyperfine energy levels of the state X$^2\Sigma^{+}(N=1)$ for vibrational states $v=0$ and $v=1$.
}
\begin{center}
\begin{tabular}{ccc}
\hline
\vspace{-6 pt}\\
State $(j,F)$&$\:v=0\,(MHz)\:$&$\:v=1\,(MHz)\:$\\
\vspace{-6 pt}\\
\hline
\vspace{-6 pt}\\
$(3/2,1)$&$192.1$&$205.8$\\
$(3/2,2)$&$161.2$&$176.5$\\
$(1/2,0)$&$155.7$&$151.9$\\
$(1/2,1)$&$0$&$0$\\
\vspace{-8 pt}\\
\vspace{-8 pt}\\
\hline
\end{tabular}
\end{center}
\end{table}

Figure \ref{fig:levels} shows a simplified version of the levels involved in laser cooling YbF, together with the wavelengths of the lasers required to excite the three X$^{2}\Sigma^{+}(N=1)$ states having $v=0,1,2$.  In reality, each of these
states has four hyperfine levels, labelled by $((N,s)j,I)F$, where $s=\frac{1}{2}$ represents the electron spin, $j$ is the intermediate quantum number for $\vec{N}+\vec{s}$, $I=\frac{1}{2}$ represents the fluorine nuclear spin, and $F$ the total angular momentum. There is no Yb nuclear spin because we use the even isotope $^{174}$Yb. Table\,\ref{tab:hfs} shows the hyperfine energies of the $v=0$ and $v=1$ manifolds \cite{Sauer96}. The $v=2$ hyperfine intervals have not been measured, but they will not differ greatly from those in $v=0$ and $v=1$. Because these intervals are large compared with the 6 MHz natural width of the A$^2\Pi_{1/2} - \mbox{X}^2\Sigma^{+}$ transition, single-frequency excitation of one hyperfine level simply pumps the molecules into the other levels and the light scattering quickly ceases. In order to prevent this, the lasers are modulated to produce sidebands that excite all four hyperfine levels. The hyperfine splitting of the upper state A$^2\Pi_{1/2}(v=0)$ is small enough \cite{Sauer99} to neglect. While the light contains all the necessary frequencies, it still does not contain all polarisations and therefore the molecules optically pump into a dark superposition of Zeeman sublevels of the X-state. This problem is overcome by applying a magnetic field at an angle to the linear polarisation of the light, which continuously mixes the bright and dark states at a rate comparable with the optical pumping rate.

We have modelled the laser cooling so that a realistic estimate can be made of the number of molecules in the fountain. The starting point is to determine the scattering rate of a molecule in the presence of the cooling light. For the moment we neglect decay to vibrational levels above $v=2$; we shall return to this issue later. For simplicity, we assume that all the hyperfine levels are excited with a single common detuning $\delta$ and that the total intensity of the light, $12I$, is equally distributed over all twelve frequency components (4 hyperfine levels times 3 vibrational states). To estimate how the population evolves in time over the 36 lower magnetic sub-levels, and the 4 upper sub-levels, we solve the rate equations for the 40-level molecule interacting with the 12 laser frequencies, including an appropriate damping of population differences between pairs of states that are mixed by the magnetic field, $B$. This damping rate is only non-zero between states of the same $j$ and $F$, that differ in $M_F$ by $\pm 1$. Between these, the damping-rate is set to $g \mu_B B$ where the g-factors are -1/3, 5/6, and 1/2 for the $(j,F) = (1/2,1)$, $(3/2, 1)$ and $(3/2, 2)$ states respectively. The branching ratios from all upper levels to all lower levels are calculated using the appendix of \cite{Wall06} together with the Franck-Condon factors $f_{0-0}=0.928$, $f_{0-1}=0.069$, and $f_{0-2}=0.0027$, taken from \cite{Zhuang11}.

Recall that the scattering rate for a 2-level system is given by
$\frac{I/I_{\rm{sat}}}{1+I/I_{\rm{sat}}+4(\delta/\Gamma)^{2}}\frac{1}{2}\Gamma$ where $\Gamma$ is the
spontaneous emission rate of the upper state and $I_{\rm{sat}}=\pi h c \Gamma/(3 \lambda^3)$ is the saturation intensity ($4.4\,$mW/cm$^2$ for our transition). Despite the complexity of the real 40-level system, one finds that the scattering rate R given by the solution to the rate equations is exceedingly well described over a wide range of intensity and detuning by a function of the same simple form:
\begin{equation}
\label{eq:rate}
R=\frac{I/(5I_{\rm{sat}})}{1+I/(5I_{\rm{sat}})+4(\delta/\Gamma)^{2}}\frac{1}{10.5}\Gamma\,,
\end{equation}
where $I$ is the laser intensity in each of the 12 frequency components. The width of the scattering resonance is still $\Gamma$ but the limiting rate at high intensity is reduced to $\Gamma/10.5$, rather than the $\Gamma/2$ of a two-level system. This is close to what one might expect -- for strong saturation the population is distributed equally over all 40 levels, leaving 1/10 of the population in the A-state, rather than the 1/2 of the two-level system. The small difference between 10 and 10.5 is probably due to the dark states which slow down the scattering a little while they are rotated into bright states by the $0.3\,$mT magnetic field. The 5-fold increase in the effective saturation intensity similarly reflects the higher multiplicity of the ground state compared with the upper state. See Appendix A for further discussion of this.

The simple form of (\ref{eq:rate}) allows a straightforward extension of standard Doppler cooling theory \cite{Metcalf99} to our case, where the saturation parameter $s=I/I_{\rm{sat}}$ is replaced by $s_{\rm{eff}}=I/(5I_{\rm{sat}})$
and $\Gamma$ is replaced, where appropriate, by $\Gamma_{\rm{eff}}=\frac{2}{10.5}\Gamma$. In a 1D molasses, the coefficient of the frictional force $-\alpha v$ is then
\begin{equation}
\label{eq:alpha}
\alpha=\frac{8s_{\rm{eff}}}{\left(1+s_{\rm{eff}}+4(\delta/\Gamma)^2\right)^{2}}\frac{\hbar \,k^{2}\delta\,\Gamma_{\rm{eff}}}{\Gamma^2}\,,
\end{equation}
and the minimum temperature,
\begin{equation}
\label{eq:Tmin}
T_{\rm{min}}=\frac{\hbar\Gamma}{2k_B}\sqrt{1+s_{\rm{eff}}}\,,
\end{equation}
is reached for a detuning of
\begin{equation}
\label{eq:deltamin}
\delta_{\rm{Tmin}}=-\frac{\Gamma}{2}\sqrt{1+s_{\rm{eff}}}\,.
\end{equation}

These formulae provide analytical insight into the operation of the molasses, but cannot tell us the number of ultracold molecules it produces. To determine this we have done a three dimensional numerical simulation of molecules moving through the 6-beam molasses. We use the effective excitation rate $\frac{\Gamma_{\rm{eff}}}{2} \frac{s_{\rm{eff}}}{1+(2\delta/\Gamma)^{2}}$, and the effective decay rate $\Gamma_{\rm{eff}}$, so as to reproduce the scattering rate given by equation (\ref{eq:rate}). The detuning $\delta$ includes the Doppler shift $v_i/\lambda$ where $v_i$ is the velocity component in the direction of beam $i$. The simulation tracks the position and velocity of each molecule by taking account of the momentum it exchanges with the light whenever a photon is absorbed or emitted.

\begin{figure}
\begin{center}
\includegraphics[width=\linewidth]{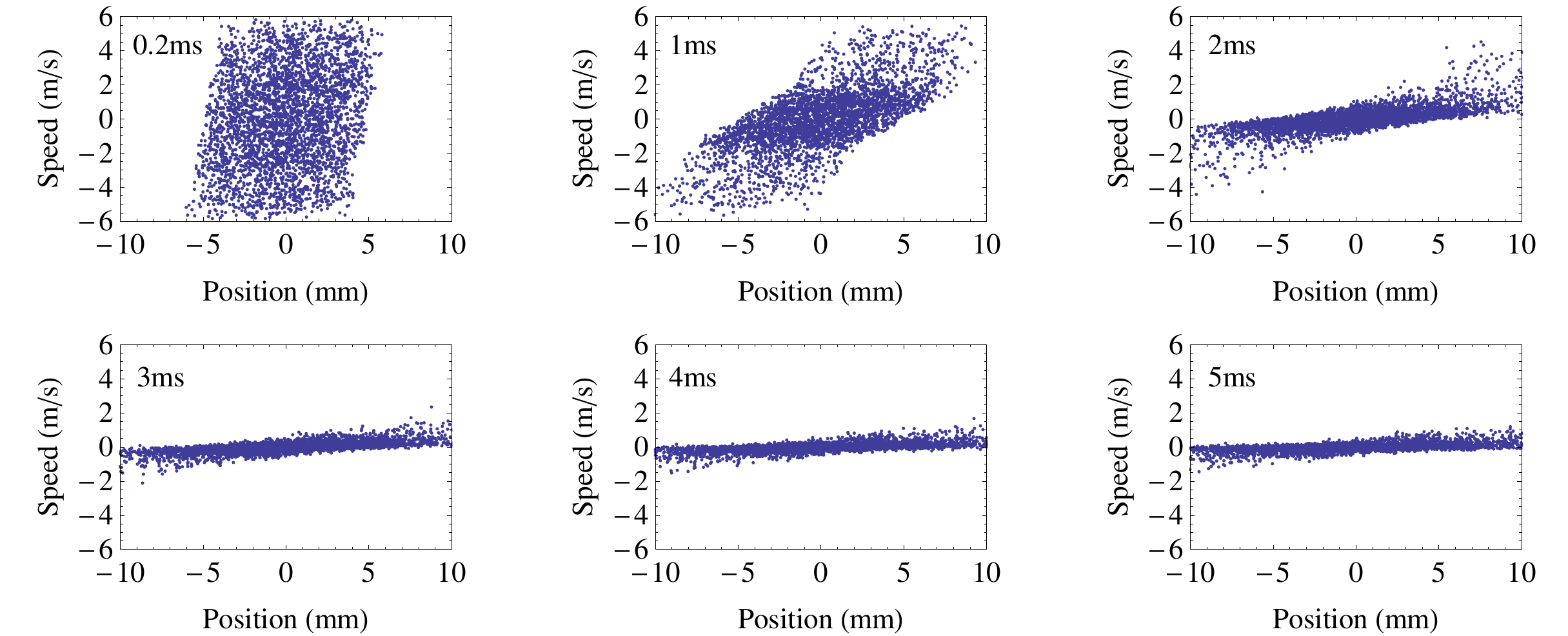}
\caption{\label{fig:psd}Phase space plots showing the distribution of molecules in a
plane transverse to the guide axis as they are cooled by the optical molasses.
}
\end{center}
\end{figure}

The molecules emerge from the $10\,\mbox{mm}$-diameter guide described in Sec.\,\ref{sec:Guide} at a temperature of 3.6\,K, with the transverse velocity truncated at $\pm 5.9\,$m/s. The molasses is centred $1.3\,\mbox{cm}$ downstream from the end of the permanent magnets. Each light beam has an intensity profile $e^{-r^{2}/w^{2}}$ with $w=1\,$cm and each laser has a total power of $275\,$mW. The laser power is recycled to form the 6 beams. The detuning from resonance is $\delta=-2\pi\times5\,$MHz. We first simulate a set of molecules that have the full distribution of velocities but all exit the guide at the same moment in time. Figure \ref{fig:psd} shows the distribution of molecules projected onto the $x-v_x$ plane of phase space at several moments of time since they emerge from the guide ($x$ is the vertical direction in figure \ref{fig:overview}). Initially, the acceptance of the guide is filled and the phase space density is almost uniform. After $5\,$ms, the transverse velocity is strongly compressed corresponding to ultra-low temperature. Similar results are found in the other two directions. From the simulation we find that the molasses has a capture velocity of approximately 10\,m/s.

\begin{figure}
\begin{center}
\includegraphics[width=0.6\linewidth]{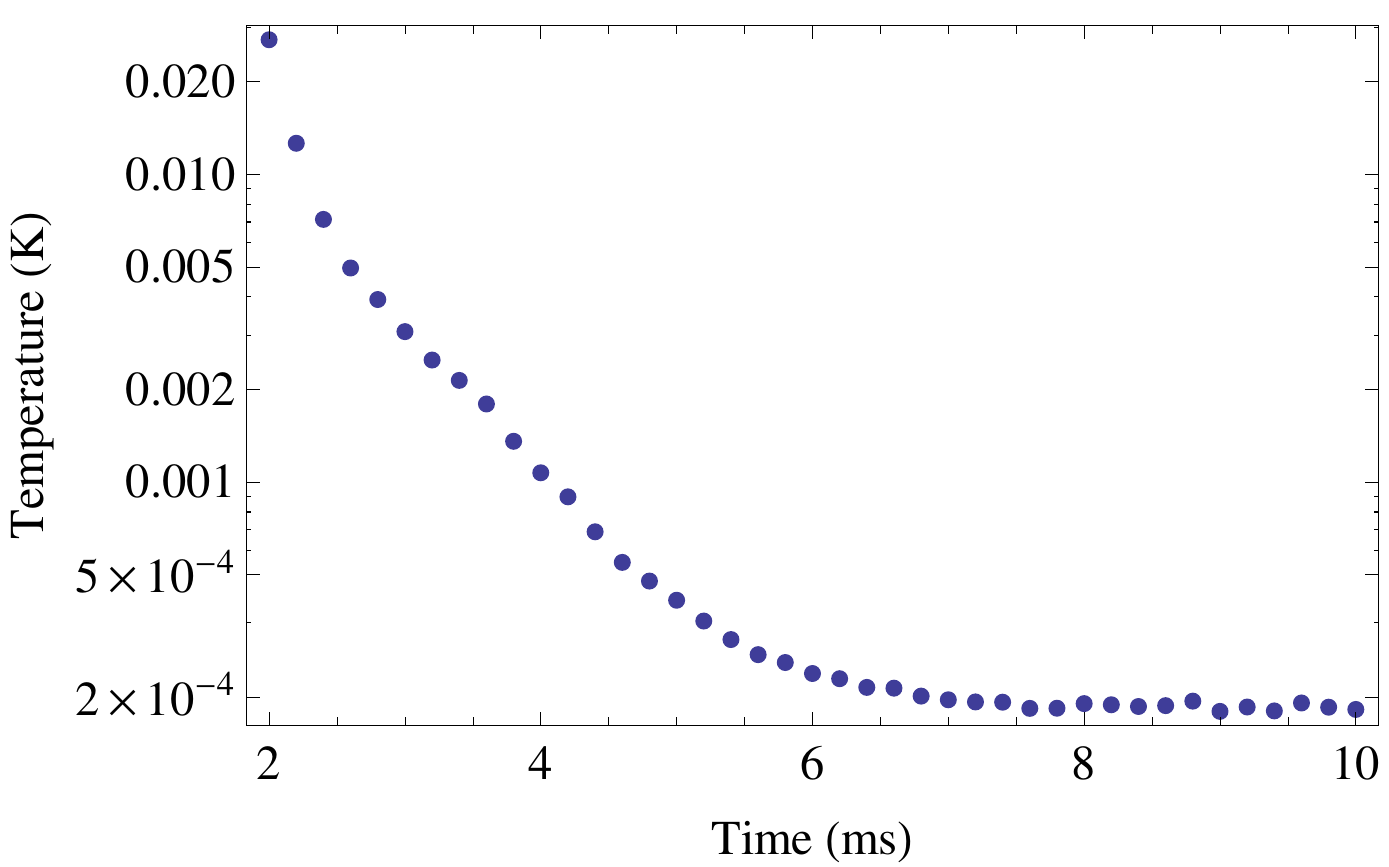}
\caption{\label{fig:temp} Temperature of the molecules versus the time since they exit the guide.
}
\end{center}
\end{figure}

Once the molecules enter the molasses a fraction of them begin to cool to low temperature. Figure \ref{fig:temp} shows how the temperature of this cold fraction decreases over time, as determined by fitting their velocity distribution to the Boltzmann distribution. We see that the temperature reduces to $185\,\mu$K, close to the Doppler limit of $194\,\mu$K given by equation (\ref{eq:Tmin}) for the parameters used in the simulation. It takes about 7\,ms for the molecules to reach this base temperature. It is likely that the temperature will actually go lower than that because of additional cooling by the Sisyphus mechanism \cite{Metcalf99}, which is not included in our simulation. Conservatively, we assume for the moment that the temperature is $185\,\mu$K.

We turn now to the probability of decay into the $v=3$ ground state.  According to the simulation, the mean scattering rate in the molasses is about $1.5 \times 10^6$ photons per second. Taking the probability of decay into $v=3$ as $3\times 10^{-4}$, slightly below the upper limit determined in \cite{Zhuang11}, we see that molecules are pumped into $v=3$ in only $2\,$ms - less than the $7\,$ms it takes to cool them. For that reason, an additional  laser returns the $v=3$ population to the cooling cycle by spontaneous decay via the $v=1$ $A$-state, as indicated in figure \ref{fig:levels}. This repumping can be much faster than $2\,$ms, allowing the full scattering rate for cooling to be maintained for much longer. Eventually, however, the molecules will decay into $v=4$, so a balance needs to be struck; the $v=3$ repumping rate needs to be high enough that the molecules are adequately cooled, but not so high that the population is pumped into $v=4$.

The molecules we aim to capture are those with forward speeds between $2\,$m/s and $10\,$m/s, which emerge from the guide over an interval of approximately $100\,$ms. The molasses must  therefore be applied for that long. We estimate that the branching ratio to $v>3$ is approximately $(0.07)^{4} = 2.4 \times 10^{-5}$, based on the trend of the Franck-Condon factors measured in \cite{Zhuang11}. If the molasses were fully on for $100\,$ms,  $70\%$ of the molecules captured would subsequently be lost to the higher vibrational states. This is avoided by repumping more slowly so that the molecules spend some time in the $v=3$ state, where they are not excited by the cooling lasers. 

We have simulated this more complicated cooling situation. The molecules entering the guide are assumed to have a time distribution $t^{2} e^{-t^{2}/(2 \sigma_{t}^{2})}$ with $\sigma_{t}=10$\,ms. Their distribution of arrival times at the exit of the guide is determined from this distribution and the distribution of forward speeds. They enter the molasses where the parameters are the same as before, except that now a very weak $v=3$ repump laser is added. For each molecule the simulation proceeds as follows. The molecule scatters $n$ photons before decaying to $v=3$, where $n$ is drawn at random according to the geometric probability distribution $r_{3}(1-r_{3})^n$ with $r_{3}=3\times 10^{-4}$. It remains in $v=3$ until, after a time $T$, it is repumped into the cooling cycle. This time is chosen at random from the exponential distribution $e^{-T/\tau_{3}}$, where we set the characteristic repumping time to $\tau_{3}=8$\,ms. This cycle continues until the molasses is turned off. A molecule is lost to $v=4$ if it has scattered more than $m$ photons in total, where $m$ is drawn at random from the geometric probability distribution $r_{4}(1-r_{4})^m$ having $r_{4}=2.4\times 10^{-5}$. Of all the molecules that are cooled, 35\% are lost in this way. There is also loss through spontaneous vibrational decay within the X state. This changes the rotational state from $N=1$ to $N=0$ or $2$, causing the molecule to decouple from the cooling cycle. The lifetimes of the excited vibrational states are given in figure \ref{fig:levels}. We calculated these using the dipole gradient $d\mu/dR = 59$\,Debye/nm \cite{Dolg92}. We find that a further 19\% of all the cooled molecules are lost this way.

\begin{figure}
\begin{center}
\includegraphics[width=0.6\linewidth]{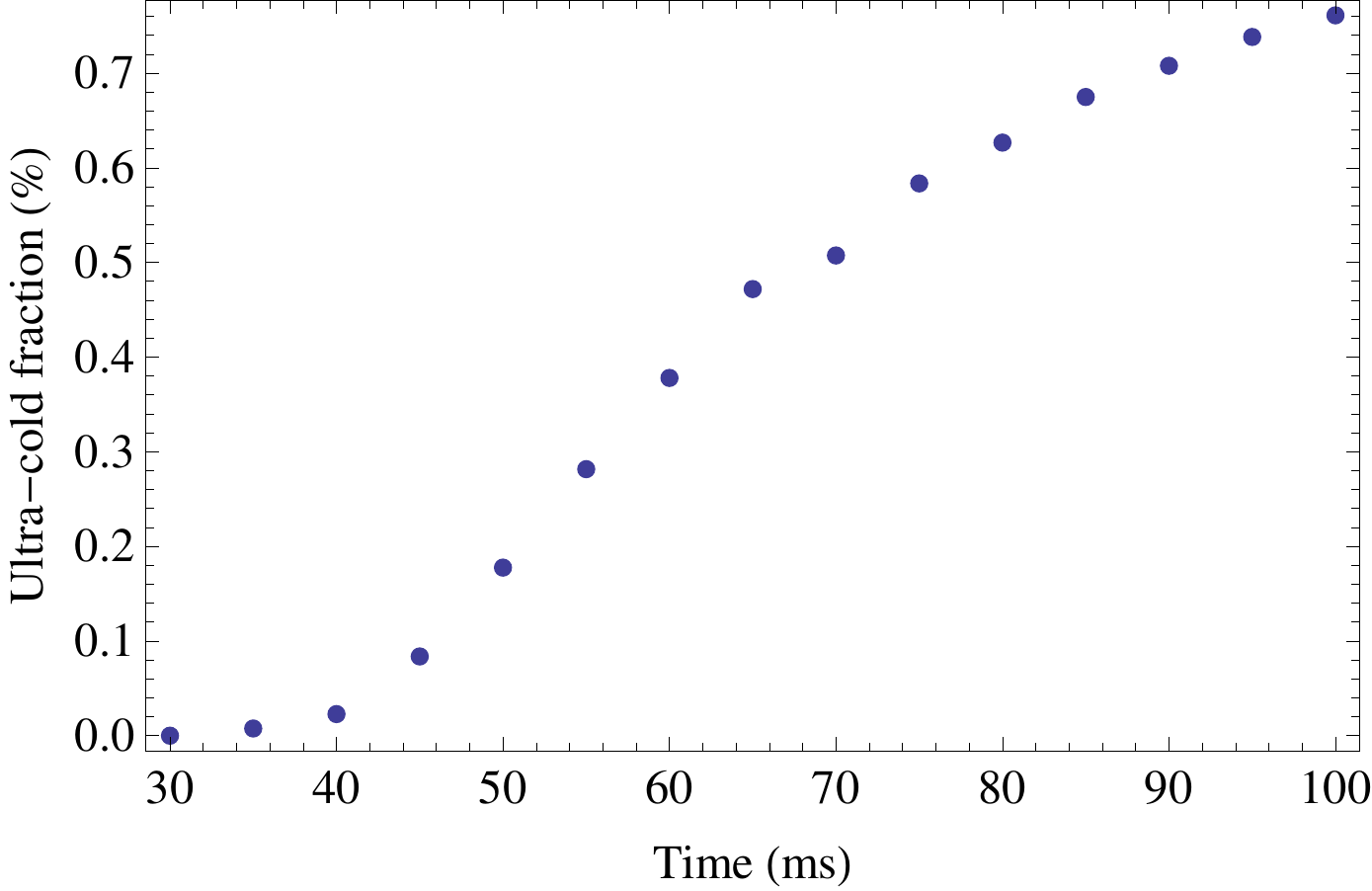}
\caption{\label{fig:number} Fraction of ultra-cold molecules versus time since the molecules were produced in the source. Here ultra-cold means that $v\le3$, that the speed is smaller than 30\,cm/s, and that the molecule is within 2\,cm of the centre of the molasses.
}
\end{center}
\end{figure}

Figure \ref{fig:number} shows the fraction of molecules cooled to ultra-cold temperature as a function of the time since they were produced. This is given as a percentage of all the molecules exiting the guide. The ``ultra-cold'' molecules are taken as those having $v\le3$, a total speed less than $30\,$cm/s and a displacement of less than 2\,cm from the centre of the molasses. Losses due to decay between the X state vibrational levels are included. Molecules begin to load into the molasses after about 30\,ms when the fastest ones that can be captured start to arrive. This ultra-cold fraction increases to $0.76\%$ at $t=100$\,ms, when the molasses is turned off. Holding the molasses on for longer does not substantially increase this fraction. The temperature of the ultra-cold distribution is 185\,$\mu$K and the spatial distribution has an rms width of approximately 0.8\,cm in each direction.

Because the $v=3$ repumping is so slow, the required laser intensity  is low: we estimate  $\sim30\,\mu$W/cm$^{2}$ on each of the four hyperfine components. Instead of adding rf sidebands to address each component, a broadband laser could be used, or a single frequency laser could be scanned back and forth over the hyperfine structure. These options make the $v=3$ repumping particularly simple to achieve. 

\section{\label{sec:Fountain} Molecular Fountain}

Once the molecules are cooled, moving molasses provides a convenient way to launch them. The upward- and downward-going laser beams are detuned by a frequency  $\pm v/\lambda$  so that the rest frame of the molasses moves upward at a velocity of $v$ (where the Doppler shift cancels the frequency difference).  Next, the beams that excite the state $(v,j,F)=(0,\frac{1}{2},1)$ are switched off and replaced by a single retro-reflected beam, circularly polarised along the magnetic field. This excites the Zeeman sublevels $m_{F}=-1,0$ but not $m_{F}=+1$, which becomes dark. All the molecules are optically pumped into this dark state.  All the light is then turned off, leaving the cloud to fly freely upward. With a velocity of $1.5\,$m/s, the molecules reach a height  of 11\,cm before falling down again under gravity, returning to the optical interaction region after 300\,ms.
The cloud expands as it flies, so not all the molecules return through an aperture at the bottom of the trajectory. With a $1\mbox{cm}\times4\mbox{cm}$ aperture, formed by the electrodes of the EDM experiment, the Maxwell-Boltzmann distribution gives a transmission of 7.5\% for a temperature of $185\,\mu$K and a flight time of $300\,$ms. This fraction of the cloud returns to the optical molasses region, where it is detected by laser-induced fluorescence using low-intensity optical molasses. Because each molecule is able to scatter thousands of photons, they are all detected. This constitutes the molecular fountain that will be used to measure the electron EDM.

Table \ref{tab:accounts} collects together the various factors discussed above that determine the number of cold molecules available for the electron EDM measurement. This shows that the fountain can provide $4.4\times 10^{5}$ detected molecules per beam pulse in the state X$^{2}\Sigma^{+}(v=0,N=1)$ with $(j,F,m_{F})=(\frac{1}{2},1,+1)$.

\section{\label{sec:Measurement} Electron EDM measurement}

\begin{figure}[tb]
\begin{center}
\includegraphics[scale=1]{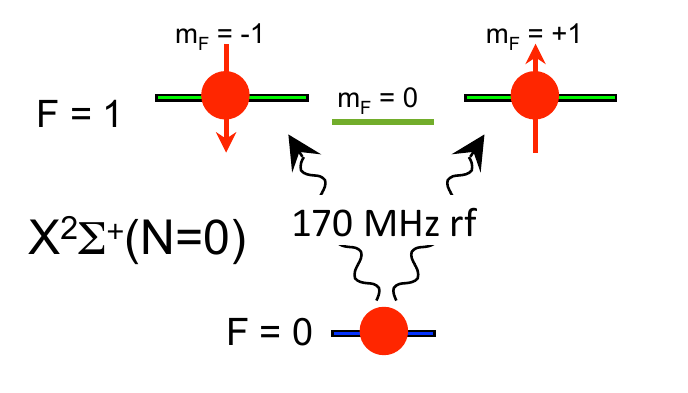}
\caption{\label{fig:hfs}Hyperfine levels of the YbF ground state X$^{2}\Sigma^{+}(v=0, N=0)$ in the presence of an electric field. Molecules initially in the $F=0$ state are excited by an rf pulse to a coherent superposition of the $F=1,m_{F}=\pm1$ states.
}
\end{center}
\end{figure}

As the molecules fly up, a microwave $\pi$-pulse at $14\,$GHz drives them to the absolute ground state $(N=0,F=0)$, indicated in figure \ref{fig:hfs}. The magnetic field is then turned off while the molecular cloud flies up into the electric field plates of the EDM experiment, illustrated schematically in figure \ref{fig:overview}.  Once it is there, we raise the field from zero to $E\approx10^6\,$V/m.  This allows the cloud to enter the plates without being accelerated (to $5.5\,$m/s) and defocussed by the gradient of the fringe field, which would damage the sensitivity through the loss of molecules. The plates also form part of a TEM transmission line operating at $170\,$MHz, as demonstrated in \cite{Hudson2007}. A $\pi$-pulse of rf power transfers all the molecules to the $F=1$ hyperfine state. Because the rf magnetic field is perpendicular to $E$, which defines our $z$ axis, the molecules are driven into the superposition state $\frac{1}{\sqrt{2}}(|m_{F}=+1\rangle+|m_{F}=-1\rangle)$, as illustrated in figure \ref{fig:hfs}. This corresponds to an alignment of the spin along the $x$-axis. A weak static magnetic field B is also applied along $z$. Under the magnetic and electric dipole interactions, the spin precesses in the $xy$ plane at an angular frequency of $(\mu_{B}B+d_{e}\eta E)/\hbar$, where the first term is the normal Zeeman interaction and the second is the small electron EDM interaction of interest, magnified by the enhancement factor \cite{enhancement} $\eta$.

The molecules now fly freely for a time $\tau\sim250$\,ms, while they rise up to the top of the plates and fall down again, accumulating a total precession angle $\phi=(\mu_{B}B+d_{e}\eta E)\tau/\hbar$. (The blackbody radiation has no appreciable effect on the molecules over this amount of time \cite{Buhmann08}). A second rf $\pi$-pulse repopulates the $F=0$ state with probability $\cos^2\phi$. The static electric field is switched off and those molecules that have been returned to $F=0$ are then driven back to $N=1$ by a second microwave pulse, and detected using the optical molasses to induce fluorescence. The fluorescence signal is proportional to the number of molecules and hence to $\cos^2\phi$. A reversal of the electric field changes $\phi$ to $(\mu_{B}B-d_{e}\eta E)\tau/\hbar$, leading to a small change in the fluorescence intensity. This change is maximized by setting the magnetic field to 35 pT, so that $\phi\approx\pi/4$ (or $-\pi/4$ if $B$ is reversed).

Thus the electron EDM is measured by recording the molecular fluorescence for each beam shot, together with the signs of $E$ and $B$. After each shot, the apparatus is switched to a new configuration and the next fluorescence pulse is measured. The EDM phase $d_{e}\eta E\tau/\hbar$ is derived from many such measurements by adding or subtracting the fluorescence signals according to the sign of the product $EB$. In practice, it is very beneficial to modulate more parameters, as detailed in \cite{Kara12}, but this is the basic principle. The statistical uncertainty in $d_{e}$ is
\begin{equation}
\label{eq:sigma }
\sigma_{d}=\frac{100}{e}\frac{\hbar}{2\eta E\tau\sqrt{N}}\,,
\end{equation}
where the first factor converts from the SI units of C.m to the conventional e.cm. In the second factor, $\eta E = 1.4\times 10^{12}\,$V/m is the effective electric field interacting with the electron EDM when the applied field $E$ is $10^6\,\mbox{V/m}$ \cite{enhancement}, $\tau$ is the free precession time and $N$ is the total number of molecules detected in the experiment. With $4.4\times10^5$ molecules per shot (Table \ref{tab:accounts}) and two shots per second, the statistical noise will be $6 \times10^{-31}\,$e.cm in $8\,$hours, the level indicated by the solid arrow in figure \ref{fig:limits}.

In order to achieve sensitivity at this statistical limit, the noise due to random fluctuations of the magnetic field must be suppressed to below $25\,\mbox{fT}/ \sqrt{\mbox{Hz}}$, which can be achieved with good magnetic shielding, together with the use of appropriate materials inside the apparatus. Excess noise due to non-statistical fluctuations in molecule number can be removed by normalising to the fluorescence intensity detected in the initial molasses.

Systematic errors will also need to be under control. References\,\cite{Hudson11,Kara12}  describe all the significant systematic errors that are known, and strategies for dealing with them. A leading concern will be the magnetic field change associated with electric field reversal.  For this reason, leakage currents will be reduced to the pA level, as demonstrated in \cite{Romalis01}. Even when there is no leakage, the currents that flow when the plates are charged (about 10\,$\mu$A if the field is switched in 10\,ms) can magnetise the shields. A design to reduce this adequately is discussed in \cite{Regan01}. It will be important to show by direct measurement that the magnetic field is properly behaved. For this purpose, we plan to use fibre- coupled SERF magnetometers \cite{Kominis03}. This recently-developed device is ideally suited to our needs: it has a field sensitivity of a few $\mbox{fT}/ \sqrt{\mbox{Hz}}$ and operates at fields below $10\,$nT. Another systematic error considered in \cite{Hudson11} arises from the geometric phase \cite{Tarbutt09} due to the (slight) rotation of the electric field viewed from the rest frame of the molecules. In \cite{Hudson11} this gave an uncertainty of $3\times 10^{-30}\,\mbox{e.cm}$. In the proposed apparatus, the geometric phase will be of similar size but the coherence time will be 390 times longer, making the EDM uncertainty correspondingly smaller, and thus entirely negligible.

\section{\label{sec:Summary} Summary}
In conclusion, we have shown that laser cooling of molecules will make it possible to build an intense, cold molecular fountain - an important new tool for quantum metrology and fundamental physics studies. We have outlined how such a source may be used to improve experimental sensitivity to the EDM of the electron by up to three orders of magnitude.  This specific application is important because it probes new elementary particle physics at energy scales up to 100 TeV and promises to illuminate the CP-violating physics of the very early universe.

\ack
This work was supported by the Royal Society, the UK funding agencies STFC and EPSRC, and the European Research Council.

\appendix
\section{Further discussion of excitation by the molasses lasers}
\setcounter{section}{1}
Consider a molecule with excited state $e$ that is coupled by the laser beams of the molasses to $N$ ground states $g_j$. Let the probabilities of occupation, $n_e$ for the excited state and $n_j$ for the ground states, be governed by the rate equations
\begin{equation}
\frac{\partial\,n_{j}}{\partial t }=A_j n_e + R_j (n_e-n_j)\,,
\end{equation}
where $A_j$ is the partial spontaneous decay rate from $e$ to $g_j$ and $R_j$ is the excitation rate for that transition, together with the normalisation condition 
\begin{equation}
n_e+\sum\limits_{j=1}^N n_{j}=1.
\end{equation}
In the steady state, where $\partial n_j/\partial t=0$, these equations give an excitation probability of 
\begin{equation}
n_e=\frac{1}{(N+1)+\sum\limits_{j=1}^N A_j/R_j}.
\end{equation}
The ratio $A_j/R_j$ can be written as 
$2 I_{sat,j}/I_j$, where $I_j$ is the intensity of the light driving the $j^{th}$ transition. $I_{sat,j}=\pi h c \Gamma/(3\lambda_j^3)$ is the saturation intensity for that transition,  $\lambda_j$ being the wavelength. $\Gamma$ is the total decay rate of state $e$. This gives
\begin{equation}
n_e=\frac{1}{(N+1)+2\sum\limits_{j=1}^N I_{sat,j}/I_j}.
\end{equation}
As the wavelengths involved in the laser cooling do not vary greatly from one vibrational state to another, it is instructive to approximate the $I_{sat,j}$ as a constant $I_{sat}$, which can be brought outside the sum to give 
\begin{equation}
n_e\simeq\frac{1}{(N+1)+2I_{sat}\sum\limits_{j=1}^N 1/I_j}.
\end{equation}
In order to have the strongest cooling, one wants to make $n_e$, and hence the scattering rate, as high as possible given the available laser intensity. It is most efficient to choose equal values for all the $I_j$, making $I_j=I_{tot}/N$, where $I_{tot}$ is the total intensity of the light. Then the excited state probability becomes simply
\begin{equation}
n_e\simeq\frac{1}{(N+1)+2N^{2}I_{sat}/I_{tot}}\,,
\label{excited fraction}
\end{equation}
and the ground state populations $n_j$ are all equal to $(1-n_e)/N$. When $I_{tot}<<\frac{2N^2}{N+1}I_{sat}$, the intensity is low and  $n_e$ tends to $(1/N^2) I_{tot}/(2 I_{sat})$. At the opposite extreme, when the intensity is high, the population is shared equally among all the levels and $n_e$ tends to $1/(N+1)$.

In the case of YbF there are 36  lower states, these being the twelve sublevels of $F=2,1,1,0$ multiplied by the $3$ vibrational states $v=0,1,2$. Instead of one upper level, there are four: the sublevels of $F=0,1$. At any given moment, each excited state is only coupled to a subset of the ground states, those having a non-vanishing transition matrix element. However, we apply a magnetic field that mixes light and dark states so that every state can be excited. Our numerical simulation yields an empirical form for the scattering rate, given in equation (\ref{eq:rate}).  After dividing by spontaneous emission rate $\Gamma$, setting detuning $\delta$ to zero, and writing $I_{tot}=12I$, this equation gives the empirical excited state probability as 
\begin{equation}
n_e\simeq\frac{4}{42+2\times35.5^{2}I_{sat}/I_{tot}}\,.
\end{equation}
Comparison with equation  (\ref{excited fraction})  shows that, despite the complexity of the real system, it is not very different in the end from four upper levels, each separately connected to 36 lower levels.


\section*{References}
\begin{thebibliography}{00}
\bibitem{Lebedev02}
Lebedev O and Pospelov M 2002 {\it Phys. Rev. Lett.} {\bf 89}, 101801
\bibitem{CKM73}
Kobayashi M and Maskawa T 1973 {\it Prog. Theor. Phys.} {\bf 49} 652
\bibitem{Khriplovich91}
Khriplovich I B and Pospelov M E 1991 {\it Sov. J. Nucl. Phys.} {\bf 53} 638
\bibitem{Sakharov67}
Sakharov A D 1967 {\it JETP} {\bf 5} 24; republished as {\it Sov. Phys. Uspekhi} {\bf 34} 392
\bibitem{Pospelov05}
Pospelov M and Ritz A 2005 {\it Ann. Phys.} {\bf 318} 119
\bibitem{Khriplovich97}
Khriplovich I B and Lamoreaux S K 1997 {\it CP Violation Without Strangeness} (Springer, New York)
\bibitem{Commins99}
Commins E D 1995 {\it Advances in AMO Physics} {\bf 40} (Academic Press, New York), pp. 1-55
\bibitem{Sandars64}
Sandars P G H and Lipworth E 1964 {\it Phys. Rev. Lett.} {\bf 13} 718
\bibitem{Schiff63}
Schiff L I 1963 {\it Phys. Rev.} {\bf 132} 2194
\bibitem{Sandars65}
Sandars P G H 1965 {\it Phys. Lett.} {\bf 14} 194
\bibitem{Regan02}
Regan B C, Commins E D, Schmidt C J and DeMille D 2002 {\it Phys. Rev. Lett} {\bf 88} 071805
\bibitem{Hinds97}
E. A. Hinds 1997 {\it Physica Scripta} {\bf T70} 34
\bibitem{enhancement}
Kozlov M G and Ezhov V F 1994 {\it Phys. Rev. A} {\bf 49} 4502; Titov M A, Mosyagin M, Ezhov V 1996 {\it Phys. Rev. Lett.} {\bf 77} 5346; Kozlov G 1997 {\it J. Phys. B} {\bf 30} L607;  Quiney H M, Skaane H, Grant I P 1998 {\it J. Phys. B} {\bf 31} L85; Parpia F A 1998 {\it J. Phys. B} {\bf 31} 1409; Mosyagin N, Kozlov M, Titov A 1998 {\it J. Phys. B} {\bf 31} L763; Nayak M K and Chaudhury R K 2006 {\it Chem. Phys. Lett.} {\bf 419} 191.  These seven calculations give the effective electric field along the internuclear axis of the YbF molecule as $(3.1,1.9,2.6,2.6,2.6,2.5,2.3)$\,TV/m respectively (after correcting a trivial factor of two error in Quiney \textit{et al.}). We take the value to be $2.5\,$TV/m. With the molecule in a $1\,$MV/m external field, the projection of the internuclear axis onto the external field reduces this to $1.4\,$TV/m. The $7.2\,$GV/m effective field in the Tl experiment\cite{Regan02} was 200 times smaller.
 \bibitem{Hudson02}
Hudson J J, Sauer B E, Tarbutt M R and Hinds E A 2002 {\it Phys. Rev. Lett.} {\bf 89} 023003
 \bibitem{Hudson11}
Hudson J J, Kara D M, Smallman I J, Sauer B E, Tarbutt M R and Hinds E A 2011 {\it Nature} {\bf 473} 493
\bibitem{Commins10}
Commins E D and DeMille D 2010 {\it Lepton Dipole Moments} eds Roberts and Marciano (World Scientific, Singapore) ch. 14
 \bibitem{DeMille02}
DeMille D 2002 {\it Phys. Rev. Lett.} {\bf 88} 067901
\bibitem{Andre06}
Andr\'e et al. 2006 {\it Nature Phys} {\bf 2} 636
\bibitem{Goral02}
Goral K, Santos L and Lewenstein M 2002 {\it Phys. Rev. Lett.} {\bf 88} 170406
\bibitem{Micheli06}
Micheli A, Brennen G and Zoller P 2006 {\it Nat. Phys.} {\bf 2} 341
\bibitem{Balakrishnan01}
Balakrishnan N and Dalgarno A 2001 {\it Chem. Phys. Lett.} {\bf 341} 652
 \bibitem{Krems08}
Krems R V 2008 {\it Phys. Chem. Chem. Phys.} {\bf 10} 4079
 \bibitem{DeMille08}
DeMille D, Cahn S, Murphree D, Rahmlow D and Kozlov M 2008 {\it Phys. Rev. Lett.} {\bf 100} 023003
\bibitem{Flambaum07}
Flambaum V V and Kozlov M G 2007 {\it Phys. Rev. Lett.} {\bf 99} 150801
\bibitem{Miranda11}
de Miranda M H G, Chotia A, Neyenhuis B, Wang D, Quemener G, Ospelkaus S, Bohn J L, Ye J and Jin D S, {\it Nat. Phys.} {\bf 7} 502
\bibitem{Danzl10}
Danzl J G, Mark M J, Haller E, Gustavsson M, Hart R, Aldegunde J, Hutson J M and N\"agerl H -C 2010 {\it Nat. Phys.} {\bf 6}, 265
\bibitem{Krems09}
Krems, Stwalley and Friedrich (eds) 2009 {\it Cold Molecules} (CRC Press)
\bibitem{Carvalho99}
deCarvalho R et al. 1999 {\it Eur. Phys. J. D} {\bf 7} 289
\bibitem{Lara06}
Lara M et al. 2006 {\it Phys. Rev. Lett.} {\bf 97} 183201
\bibitem{Soldan09}
Sold\'an P, \.Zuchowski P S and Hutson J M 2009 {\it Faraday Discussion} {\bf 142} 191
\bibitem{Wallis09}
Wallis A O G and Hutson J M 2009 {\it Phys. Rev. Lett.} {\bf 103} 183201
\bibitem{Tokunaga11}
Tokunaga S K et al. 2011 {\it Eur. Phys. J D} {\bf 65} 141
\bibitem{DiRosa04}
DiRosa M D 2004 {\it Eur. Phys. J. D} {\bf 31} 395
\bibitem{Shuman10}
Shuman E S et al. 2010 {\it Nature} {\bf 467}, 820
\bibitem{Barry11}
Barry J F, Shuman E S, Norrgard E B and DeMille D 2012 {\it Phys. Rev. Lett.} {\bf 108} 103002
\bibitem{Zhuang11}
Zhuang X, Le A, Steimle T C, Bulleid N E, Smallman I J, Hendricks R J, Skoff S M, Hudson J J, Sauer B E, Hinds E A and Tarbutt M R 2011 {\it Phys. Chem. Chem. Phys.} {\bf 13} 19013
\bibitem{Metcalf99}
Metcalf H J and Van der Straten P 1999 {\it Laser cooling and Trapping} (Springer)
\bibitem{Lu11}
Lu H - I, Rasmussen J, Wright M, Patterson D and Doyle J M 2011 {\it Phys. Chem. Chem. Phys.} {\bf 13} 18986
\bibitem{Skoff09}
Skoff S M, Hendricks R J, Sinclair C D J, Tarbutt M R, Hudson J J, Segal D M, Sauer B E and Hinds E A 2009 {\it New J. Phys.} {\bf 11} 123026
\bibitem{Skoff11}
Skoff S M, Hendricks R J, Sinclair C D J, Hudson J J, Segal D M, Sauer B E, Hinds E A and Tarbutt M R 2011 {\it Phys. Rev. A} {\bf 83} 023418
\bibitem{SkoffThesis}
Skoff S 2011 {\it PhD thesis}. http://www3.imperial.ac.uk/ccm/publications/extra
\bibitem{Patterson07}
Patterson D and Doyle J M 2007 {\it J. Chem Phys.} {\bf 126} 154307
\bibitem{Sauer96}
Sauer B E, Wang J and Hinds E A 1996 {\it J. Chem. Phys.} {\bf 105} 7412
\bibitem{Sauer99}
Sauer B E, Cahn S B, Kozlov M G, Redgrave G D, Hinds E A 1999 {\it J. Chem. Phys.} {\bf 110} 8424
\bibitem{Wall06}
Wall T E, Kanem J F, Hudson J J, Sauer B E, Cho D, Boshier M G, Hinds E A and Tarbutt M R 2008 {\it Phys. Rev. A} {\bf 78} 062509
\bibitem{Hudson2007}
Hudson J J, Ashworth H T, Kara D M, Tarbutt M R, Sauer B E and Hinds E A 2007 {\it Phys. Rev. A.} {\bf 76} 033410
\bibitem{Dolg92}
Dolg M, Stoll H and Preuss H 1992, {\it Chem. Phys.} {\bf 165} 21
\bibitem{Buhmann08}
Buhmann S Y, Tarbutt M R, Scheel S and Hinds E A 2008 {\it Phys. Rev. A} {\bf 78} 052901
\bibitem{Kara12}
Kara D M, Smallman I J, Hudson J J, Sauer B E, Tarbutt M R and Hinds E A 2012 {\it New J. Phys.} {\bf 14} 103051
\bibitem{Romalis01}
Romalis M V et al. 2001 {\it Phys. Rev. Lett.} {\bf 86} 2505
\bibitem{Regan01}
Regan B C 2001 {\it Ph.D. thesis} Physics Department, University of California, Berkeley
\bibitem{Kominis03}
Kominis I K, Kornack T W, Allred J C and Romalis M V 2003 {\it Nature} {\bf 422}, 596
\bibitem{Tarbutt09}
Tarbutt M R, Hudson J J, Sauer B E and Hinds E A 2009 {\it Faraday Discussions} {\bf 142} 37

\end{thebibliography}
\end{document}